\documentclass{nature} 
\usepackage[latin1]{inputenc}
\usepackage[english]{babel}
\usepackage{enumerate}
\usepackage{color}

\usepackage{dcolumn}
\usepackage{bm}
\usepackage{mathptmx}
\usepackage{pifont}
\usepackage[latin1]{inputenc}
\usepackage{cancel}
\usepackage{amssymb}
\usepackage{amsmath}
\usepackage{upgreek}
\usepackage{xcolor}
\usepackage{graphicx}
\usepackage[normalem]{ulem}
\newcommand\bluesout{\bgroup\markoverwith{\textcolor{blue}{\rule[0.5ex]{2pt}{1.1pt}}}\ULon}
\makeatletter
\let\saved@includegraphics\includegraphics
\AtBeginDocument{\let\includegraphics\saved@includegraphics}
\renewenvironment*{figure}{\@float{figure}}{\end@float}
\makeatother
\title{Soft-mode nonlinearities away from ferroelectric phase transition}

\author{Payel Shee,$^{1,2}$ Ipek Efe,$^{3}$ Jingwen Li,$^3$ Kshitij V. Goyal,$^{1}$ Morgan Trassin,$^{3}$ and Shovon Pal$^{1,2}$}

\begin{document}
\maketitle
\begin{affiliations}
\item School of Physical Sciences, National Institute of Science Education and Research, Jatni, 752 050 Odisha, India
\item Homi Bhabha National Institute, Training School Complex, Anushakti Nagar, 400 094 Mumbai, India
\item Department of Materials, ETH Zurich, 8093 Zurich, Switzerland
\end{affiliations}


\begin{abstract}
The interplay between ionic and electronic subsystems dictates the behavior of structural phase transitions in polar dielectrics, a coupling mediated by soft optical phonon modes. In incipient ferroelectrics such as SrTiO$_3$ (STO), strong local-field effects can drive the lattice into a non-perturbative regime near the phase boundary. However, disentangling the distinct contributions of local fields from those of spontaneous macroscopic polarization remains an experimental challenge. Here, we isolate these mechanisms by probing paraelectric STO deep within its symmetric phase, where macroscopic spontaneous polarization is suppressed. Linear terahertz (THz) spectroscopy reveals that the soft mode exhibits a hybrid character, predominantly driven by electronic polarizability. Utilizing two-dimensional THz spectroscopy, we map the underlying nonlinear signals, demonstrating that the system persists in a perturbative regime characterized by robust local-field coherence. By implementing a microscopic model of coupled electronic and lattice degrees of freedom mediated by local fields, we qualitatively reproduce these multidimensional coherent signatures. Our findings highlight that while local fields are necessary to initiate non-perturbative lattice dynamics, they are insufficient on their own. This reveals that spontaneous polarization plays a deterministic role in dictating soft-mode nonlinearities in strongly correlated polar dielectrics.
\end{abstract}

\maketitle

\section*{Highlights}
\begin{itemize}
\item Paraelectric soft optical phonon identified as a hybrid excitation via THz spectroscopy.
\item 2D THz spectroscopy reveals strong nonlinear signals without any soft-mode frequency shift.
\item Persistent photon-echo signatures confirm the presence of robust local fields.
\item Our microscopic model proves that local fields alone are insufficient to drive non-perturbative behavior with moderate field strengths.
\end{itemize}

\section*{Summary}
Soft optical phonon modes drive structural phase transitions in polar dielectric materials, with their declining frequency signaling lattice instability. These modes represent a coupled excitation of ionic and electronic degrees of freedom, a relationship quantified by the Born effective charge and mediated by internal local electric fields. Local field effects heavily dictate macroscopic dielectric properties, shaping both linear and nonlinear optical responses. Because these low-energy modes operate in the meV range, they are uniquely accessible via terahertz spectroscopy. Specifically, two-dimensional terahertz (2D THz) spectroscopy provides the phase-sensitive resolution necessary to directly probe local-field-induced nonlinear dynamics and coherences. By examining the nonlinear response of a model paraelectric system, we demonstrate that despite a strong local-field contribution, the system remains firmly within the perturbative regime of light-matter interaction. These findings clarify the nonlinear dynamics of paraelectric systems by explicitly disentangling the distinct roles of local-field effects and strong spontaneous polarization.

\section*{Introduction}
Soft optical phonon modes are central to understanding structural phase transitions in polar dielectric materials. As a system approaches a phase transition, the fundamental frequency of these modes decreases toward zero, serving as a highly sensitive indicator of lattice instability~\cite{Venkataraman1979, Scott1974, Petzelt2003, Kamba2020, Barker1962}. A defining characteristic of this soft phonon oscillation is its nature as a coupled excitation, which fundamentally integrates both ionic and electronic degrees of freedom~\cite{Cohen1992}. This hybrid behavior is quantitatively captured by the Born effective charge ($Z_{\rm eff}$), which defines the coupling strength between atomic displacements and the resulting macroscopic crystal polarization -- or equivalently, the force exerted on an ion by the local microscopic electric field. This local field, representing the superposition of the externally applied field and the internal fields generated by neighboring ions and electrons, is essential for mapping microscopic polarizability to the macroscopic dielectric response~\cite{Hannay1983, Bucher1990}. Consequently, local field effects dictate equilibrium dielectric properties~\cite{Neb2026}, significantly modifying linear optical susceptibility~\cite{Maki1991} and playing an equally critical role in governing nonlinear optical responses~\cite{Pal2021}. In dielectric materials, the pronounced local field significantly influences the dynamics of the soft optical phonon mode~\cite{Folpini2017, Pal2021}. Operating at low energies (meV range), these modes can be resonantly probed via terahertz (THz) spectroscopy~\cite{Kozina2019, Subedi2014, Yang2023}. Crucially, two-dimensional (2D) THz spectroscopy enables direct interrogation of the associated nonlinear response by providing phase-sensitive access to field interactions, thereby facilitating the identification of local field-induced coherence~\cite{Woerner2013, Houver2019, Raab2019, Markmann2021, Kuehn2011, Somma2016prl, Somma2016}. Beyond dielectrics, this local-field framework defines the many-body interactions, dephasing dynamics, and multi-quantum coherences in semiconductor quantum well systems~\cite{Peyghambarian1984, Stone2009}, while also driving the linear and nonlinear optical response in dense atomic vapors~\cite{Maki1991, Gao2016, Dolgaleva12012}.

Building on this framework, previous studies on strained ferroelectric SrTiO$_3$ (STO) elucidated the microscopic origin of soft-mode nonlinearities, demonstrating that the soft mode in a ferroelectric material behaves as a hybrid of pure lattice excitations and electronic interband transitions~\cite{Pal2021}. Due to the system's proximity to a structural phase transition, even moderate terahertz (THz) fields drove the lattice into a non-perturbative regime of light-matter interaction, manifested as a pronounced blue-shift in the soft-mode frequency~\cite{Folpini2017, Katayama2012}. The observation of photon-echo signals at negative coherence times further provided a hallmark of strong local-field-induced coherence~\cite{Pal2021}. While these findings established that strong local fields substantially modify soft-mode nonlinearities in ferroelectric perovskites, they focused exclusively on systems in the ferroelectric phase, where large spontaneous polarization and phase-boundary instabilities coexist. Consequently, the individual contributions of local field effects, macroscopic polarization, and proximity to a phase transition remain inherently intertwined. This ambiguity raises a fundamental question: are local-field effects alone sufficient to drive the soft mode into the non-perturbative regime, or are a strong spontaneous polarization and immediate proximity to a structural phase transition necessary?

In this article, we address this question by investigating the nonlinear response of a paraelectric STO system operating far from the ferroelectric phase transition. Although fundamentally paraelectric, the system exhibits a minor initial polar component that attenuates with increasing film thickness, driven by a combination of compressive strain and substrate layer polarization~\cite{Efe2024}. Using terahertz time-domain spectroscopy (THz-TDS), we find that the electronic contribution to the soft mode significantly exceeds the ionic contribution, confirming the hybrid character of the excitation. Despite this substantial electronic character and a moderately strong driving field, two-dimensional terahertz (2D-THz) spectroscopy reveals that the nonlinear response remains strictly within the perturbative regime of light-matter interaction. Nonetheless, within this perturbative regime, photon-echo signals provide unambiguous signatures of a strong local field. To elucidate the microscopic origin of this behavior, we developed a theoretical model that captures the hybrid nature of the soft mode by coupling two distinct subsystems: electronic oscillators and lattice vibrations. This inter-subsystem coupling is mediated directly by the local field~\cite{Knoester1989, Schmitt-Rink1991}. The calculated nonlinear response successfully reproduces the key features of our experimental observations, establishing a consistent microscopic framework. Our findings reveal that a strong local field alone is insufficient to drive the system into the non-perturbative light-matter interaction regime; rather, a robust contribution from spontaneous polarization is required.

\section*{Results and discussion}
\begin{figure*}[t!]
	\centering
	\includegraphics[width=0.95\linewidth]{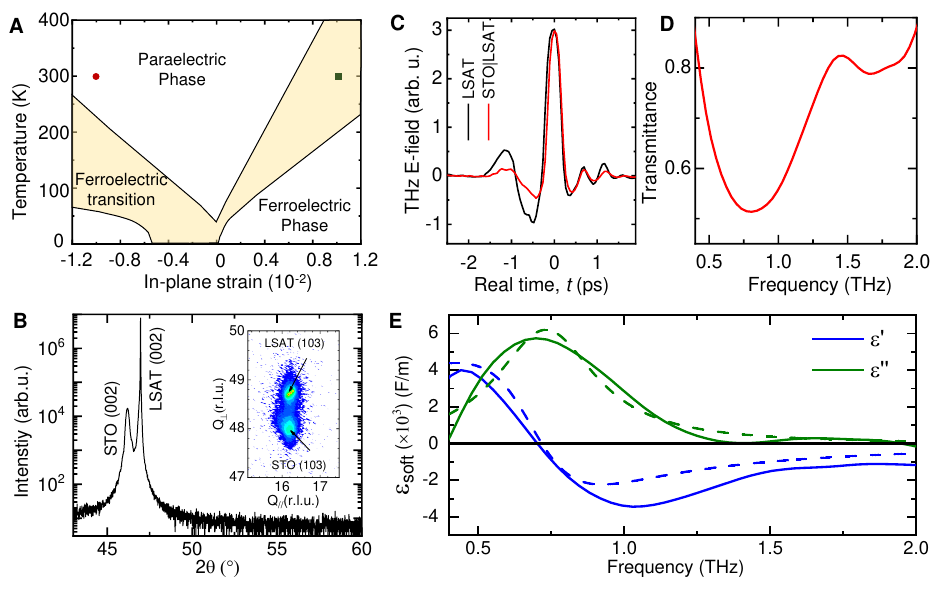}
	\caption{{\bf Structural and terahertz (THz) spectroscopic characterization of SrTiO$_3$ (STO) thin films.} {\bf (A)} Strain-phase diagram of STO derived from thermodynamic analysis (adapted from Ref.~[31]). The red circle corresponds to the paraelectric phase position of STO (current study), while the green square positions STO in the range of ferroelectric transition (Ref.~[11]). {\bf (B)} X-ray diffraction (XRD) $\theta-2\theta$ scan of a 50-nm-thick STO film grown on an LSAT substrate. The inset displays the reciprocal space map (RSM) around the LSAT(103) asymmetric reflection. {\bf (C)} Time-domain THz transients transmitted through the bare LSAT substrate (black) and the ${\rm STO}|{\rm LSAT}$ hetero-structure (red). {\bf (D)} Frequency-dependent THz transmittance exhibiting a distinct dip at approximately 0.72\,THz, indicative of the soft phonon mode. {\bf (E)} Real ($\epsilon^{\prime}$) and imaginary ($\epsilon^{\prime\prime}$) parts of the complex dielectric function. The dashed lines denote fits using a damped harmonic oscillator model, where $\epsilon^{\prime}$ crosses zero and $\epsilon^{\prime\prime}$ peaks at the soft-mode frequency.}
	\label{fig1}
\end{figure*}

A model dielectric system in the paraelectric regime was investigated, with epitaxial strain as a phase-control tuning parameter. Specifically, a 50-nm-thick (001)-oriented SrTiO$_3$ (STO) thin film was epitaxially grown on a (001)-oriented (La$_{0.3}$Sr$_{0.7}$)(Al$_{0.65}$Ta$_{0.35}$)O$_3$ (LSAT) substrate via pulsed laser deposition. The resulting lattice mismatch induces an epitaxial strain that constrains the STO layer within the room-temperature paraelectric regime, as indicated by the red dot in Figure~\ref{fig1}A~\cite{Haeni2004}. X-ray diffraction (XRD) patterns exhibit sharp, well-defined (001) reflections for both the LSAT substrate and the STO film, confirming excellent crystalline quality (Figure~\ref{fig1}B). Furthermore, reciprocal space mapping confirms coherent growth, revealing that the substrate imposes a compressive in-plane strain of 0.96\% on the STO film (Figure~\ref{fig1}B, inset). This compressive strain successfully stabilizes the paraelectric phase and tunes the soft phonon mode into the THz frequency range. Comprehensive details regarding sample fabrication and strain characterization are provided in the Supplementary Material.

Because the soft phonon mode is polar, it couples strongly with external optical fields. To determine its spectral position, we performed terahertz (THz) time-domain transmission spectroscopy at room temperature on both the bare LSAT substrate and the $\text{STO}|\text{LSAT}$ hetero-structure. The time-domain transients transmitted through the substrate and the sample are displayed in Figure~\ref{fig1}(C). The pulse transmitted through the STO thin film exhibits a distinct phase shift and attenuation compared to that of the bare LSAT. To analyze this behavior in the frequency domain, we performed a fast Fourier transform (FFT) of the time-domain signals. By normalizing the spectral response of the $\text{STO}|\text{LSAT}$ sample to that of the bare LSAT substrate, we obtained the frequency-dependent transmittance shown in Figure~\ref{fig1}(D). The spectrum reveals a broad transmission dip centered at 0.72\,THz with a full width at half maximum (FWHM) of 0.49\,THz, confirming the presence of the soft optical phonon mode in the paraelectric phase~\cite{Barker1962, Katayama2008, Sirenko2000}. Notably, this resonant frequency is slightly lower than values reported in earlier studies of strained STO systems approaching the ferroelectric transition. This frequency shift highlights the extreme sensitivity of the soft phonon dynamics to the film's epitaxial strain state.

Having identified the soft mode in the paraelectric regime, we calculate the frequency-dependent complex dielectric function within the thin-film approximation ($ d \ll \lambda_{\rm THz}$)~\cite{Tu2003}. The soft mode signature manifests clearly as a simultaneous zero-crossing in the real part and a resonant peak in the imaginary part of the dielectric function. Using Cochran's formalisms, we quantify the relative contributions of electronic and ionic responses from the linear terahertz response (see the Supplementary Material for an extended analysis)~\cite{Cochran1960}. We find that the electronic contribution to the soft mode exceeds the purely ionic counterpart by a factor of approximately five, even deep within the paraelectric phase. This finding aligns with observations reported for ferroelectric STO~\cite{Pal2021}, demonstrating that the hybrid nature of the soft mode in STO persists across both the ferroelectric and paraelectric phases. Consequently, accounting for strong electronic contributions and local-field effects resolves the apparent discrepancy regarding the enhanced effective charge ($Z_{\rm eff}$). These insights naturally raise a pivotal question: how do these pronounced electronic contributions and local-field effects dictate the nonlinear response of a paraelectric material far from its ferroelectric phase transition?
 
\begin{figure*}[t!]
	\centering
	\includegraphics[width=1.0\linewidth]{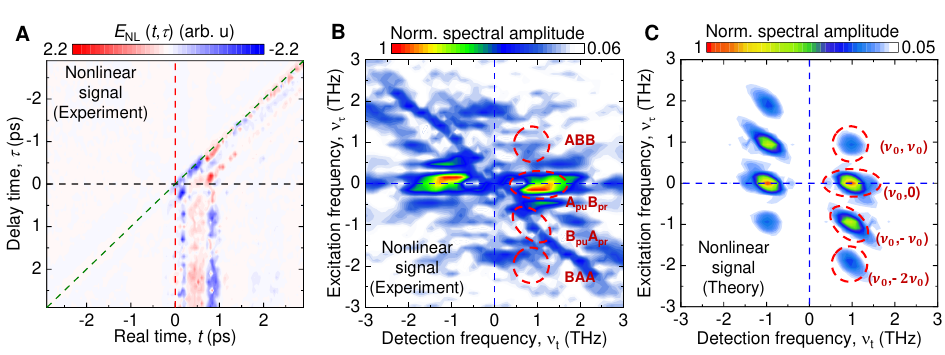}
	\caption{{\bf Experimental and theoretical two-dimensional (2D) nonlinear terahertz (THz) signals.} {\bf (A)} Emitted nonlinear THz field, $E_{\rm NL}(t,\tau)$ plotted as a function of real time $t$ and pump delay time $\tau$. Green- and red-dashed lines trace the wavefronts of the two driving THz fields, while the black-dashed line marks zero delay ($\tau = 0$). {\bf (B)} Contour plot of the experimental 2D frequency spectrum, $E_{\rm NL}(\nu_{t}, \nu_{\tau})$, obtained via a 2D Fourier transform of the time-domain signal. {\bf (C)} Simulated 2D frequency spectrum, $E_{\rm NL}(\nu_{t}, \nu_{\tau})$, derived from the theoretical model. The colored ellipses outline the spectral positions of the nonlinear signals. Blue-dashed lines in (B) and (C) represent the zero-frequency axes.}
	\label{fig2}
\end{figure*}

To this end, we performed two-dimensional THz spectroscopy employing two phase-locked THz pulses (THz-A and THz-B), separated by a delay time $\tau$. The nonlinear response from the sample is extracted as $E_{\rm NL} = E_{\rm AB}(t, \tau)-E_{\rm A}(t, \tau)-E_{\rm B}(t)$ as shown in Figure \ref{fig2}A, where $E_{\rm AB}$ denotes both pulse response, and $E_{\rm A} (E_{\rm B})$ is the single pulse response transmitted through the sample~\cite{Dutta2025, Dutta2025JPCM, Hamm2011, Elsaesser2019, Mukamel1995, Huang2026, Haldar2026}. By performing a 2D Fourier transform, we can separate all the different $\chi^{(3)}$ signals of the soft mode in the frequency domain at the detection frequency $\nu_t=\nu_0=$ 0.97\,THz, which lies within the FWHM range of the soft mode frequency. The resulting 2D spectrum shown in Figure \ref{fig2}B, reveals that the $\rm A_{\rm pu}-\rm B_{\rm pr}$ signal situated at ($\nu_t, \nu_\tau$) = (0.97, 0)\,THz is the most dominant one. In contrast, the $\rm B_{\rm pu}-\rm A_{\rm pr}$ at ($\nu_t, \nu_\tau$) = (0.97, -0.97)\,THz as well as the ABB photon echo at ($\nu_t, \nu_\tau$) = (0.97, 0.97)\,THz and BAA photon echo at ($\nu_t, \nu_\tau$) = (0.97, 1.94)\,THz are comparatively weaker in intensity.

To elucidate the microscopic origin of the observed nonlinear response, we modeled the system as two interacting subsystems that contribute to the total polarization. The first subsystem comprises electronic oscillators that represent the electronic dipoles of the Ti--O covalent bonds within the perovskite lattice. Given their intrinsic nonlinearity, these are treated as an ensemble of two-level systems. The second subsystem accounts for lattice excitations. Because the anharmonicity of the optical phonon is negligible, this subsystem is modeled as a classical damped harmonic oscillator, and its intrinsic nonlinearity is ignored. These two subsystems are self-consistently coupled through the local electric field, which combines the external THz field and the induced polarization. Consequently, the system's effective nonlinearity is primarily determined by the electronic response. The resulting equations of motion representing the two subsystems were solved in the time domain to extract the nonlinear response. A 2D Fourier transform of the time-domain data yielded the frequency-domain nonlinear signals (Figure \ref{fig2}C), which qualitatively reproduce all experimentally observed $\chi^{(3)}$ features (modeling details are elaborated in the Supplementary Information). To gain further insight into the individual signal components, we applied a 2D Gaussian spectral filter to each $(\nu_t, \nu_{\tau})$ domain feature (Figure \ref{fig2}B) and performed an inverse 2D Fourier transform to reconstruct the corresponding signals in the $(t, \tau)$ domain.

\begin{figure}[t!]
	\centering
	\includegraphics[width=0.7\columnwidth]{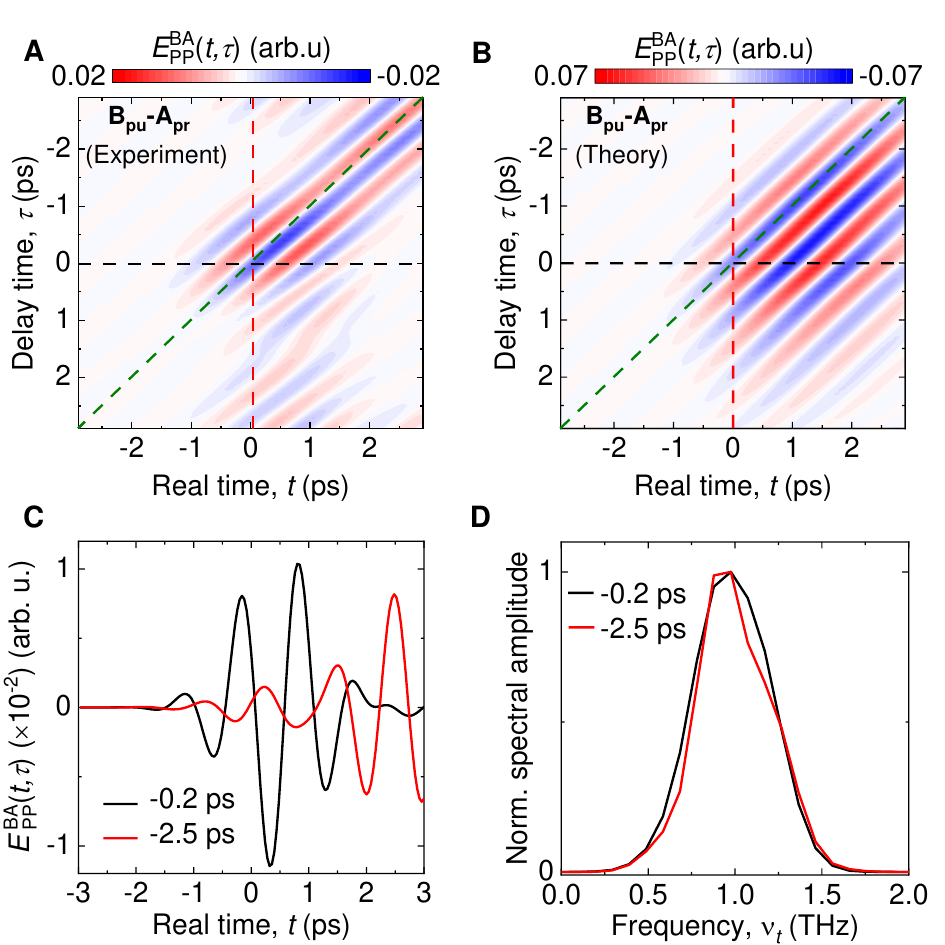}
	\caption{{\bf Temporal profiles and spectral analysis of the inverse Fourier-transformed $\rm B_{\rm pu}-\rm A_{\rm pr}$ signals.} Contour plots of the inverse Fourier-transformed pump-probe fields, $E^{\rm BA}_{\rm pp}(t,\tau)$, signals obtained from {\bf (A)} experiment and {\bf (B)} theory. Green- and red-dashed lines trace the wavefronts of the two driving THz fields, while the black-dashed line marks zero delay ($\tau = 0$). {\bf (C)} Time-domain traces of $E^{\rm BA}_{\rm pp} (t, \tau)$ at two distinct delay times. {\bf (D)} Corresponding frequency-domain spectra computed via a one-dimensional (1D) Fourier transform of the traces in (C).}
	\label{fig3}
\end{figure}

Figures~\ref{fig2}B and~\ref{fig2}C reveal two distinct pump-probe signals, designated as $\rm A_{\rm pu}-\rm B_{\rm pr}$ and $\rm B_{\rm pu}-\rm A_{\rm pr}$, which track the pump-induced dynamics via interaction with the delayed probe pulse. The experimental and theoretically modeled $\rm B_{\rm pu}-\rm A_{\rm pr}$ signals are displayed in Figures \ref{fig3}A and B, respectively. In both instances, the temporal profile of the signal closely tracks that of the probe pulse (pulse A). The strong agreement between the experimental and theoretical spectra validates the numerical model. To elucidate the nature of the pump-induced perturbation, we analyzed the temporal evolution of the $\rm B_{\rm pu}-\rm A_{\rm pr}$ signal by isolating two distinct time traces at fixed delay in Figure~\ref{fig3}C: one at $-0.2$\,ps, characterized by strong coherent pulse overlap, and another at $-2.5$\,ps, where the pulses are temporally well-separated. A one-dimensional fast Fourier transform (1D-FFT) of these transients reveals no measurable shift in the soft-mode frequency within experimental resolution. This implies that, despite its high intensity, the pump field fails to alter the soft-mode frequency. This behavior stands in stark contrast to previous observations in ferroelectric STO~\cite{Pal2021}, where comparable THz fields induced a blue shift in the soft-mode frequency -- a signature of the non-perturbative regime. Conversely, the absence of a frequency shift in the present paraelectric phase indicates that the system remains within the perturbative regime of light-matter interactions. A similar absence of a frequency shift is observed in the $\rm A_{\rm pu}-\rm B_{\rm pr}$ signal. A more rigorous framework for this behavior can be extracted from the spectrally resolved pump-probe data.

\begin{figure}[t!]
	\centering
	\includegraphics[width=0.7\columnwidth]{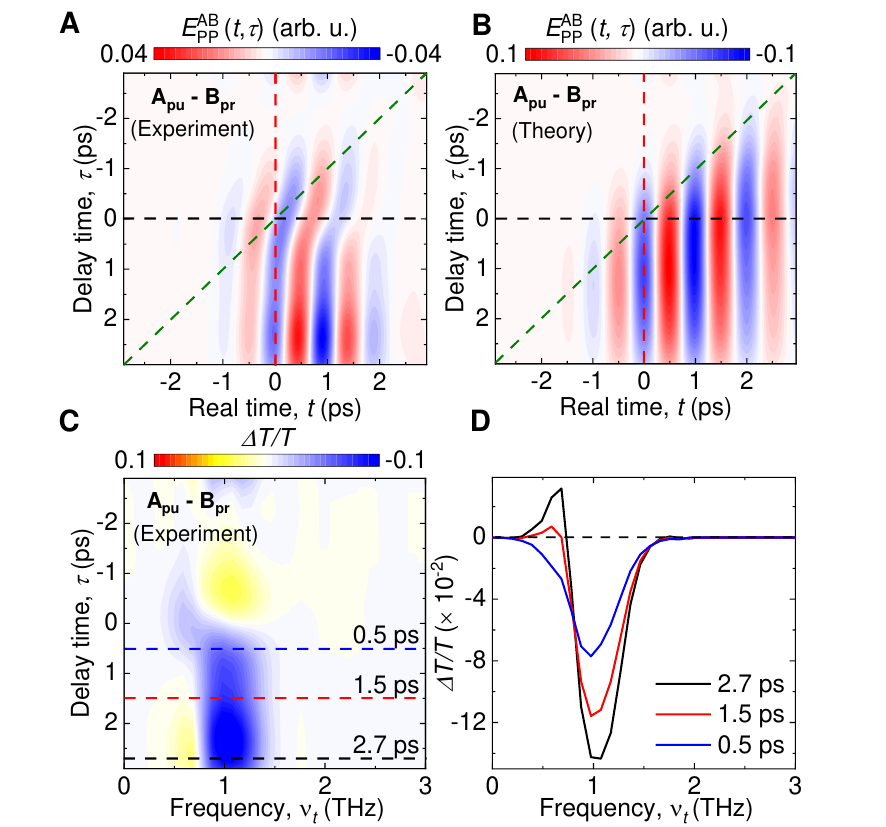}
	\caption{{\bf Temporal profiles and spectrally-resolved dynamics of the inverse Fourier-transformed $\rm A_{\rm pu}-\rm B_{\rm pr}$ signal.} Contour plots of the nonlinear pump-probe fields, $E^{\rm AB}_{\rm pp}(t,\tau)$, obtained from {\bf (A)} experiment and {\bf (B)} theory. Green- and red-dashed lines trace the wavefronts of the two driving THz fields, while the black-dashed line marks zero delay ($\tau = 0$). {\bf (C)} Experimental contour plot of the spectrally-resolved signal, $\rm S^{AB}_{pp}(\nu_t,\tau)$. {\bf (D)} Traces of $\rm S^{AB}_{pp}(\nu_\textit{t},\tau)$ at three distinct delay times.}
	\label{fig4}
\end{figure}

To investigate this further, we obtained the experimental and theoretical $\rm A_{\rm pu}-\rm B_{\rm pr}$ signals, as displayed in Figures~\ref{fig4}A and~\ref{fig4}B, respectively, which exhibit strong qualitative agreement. In this configuration, the temporal profile of the pump-probe signal follows that of the probe pulse (pulse B). The corresponding spectrally-resolved $\rm A_{\rm pu}-\rm B_{\rm pr}$ signals, $S_{\rm pp}^{\rm AB}(\nu_t, \tau)$, are presented in Figures~\ref{fig4}C and~\ref{fig6}A, evaluated using the relation: $S_{\rm pp}^{\rm AB}(\nu _{t},\tau) = 2\mathrm{Re}[E_{\rm pp}^{\rm AB}(\nu _{t},\tau )E_{\rm B}^{*}(\nu _{t})]/|E_{\rm B}(\nu _{t})|{}^{2}$, where \(E_{\rm B}^*(\nu_t)\) is the complex conjugate of $E_{\rm B}(\nu_t)$. The simulated spectrally resolved signals closely match the experimental data. Figure~\ref{fig4}D and Figure~\ref{fig6}D present temporal slices of the experimental and theoretical spectrally-resolved signals, respectively, at delay times of 0.5\,ps, 1.5\,ps, and 2.7\,ps. Both the theoretical and experimental profiles exhibit purely absorptive features at the soft-mode frequency. The complete absence of an asymmetric, dispersive-like lineshape directly confirms that no dynamic frequency shift occurs during the light-matter interaction. This finding further substantiates that the nonlinear response of paraelectric STO remains strictly confined within the perturbative regime for moderate THz pump fields.

\begin{figure}[t!]
	\centering
	\includegraphics[width=0.7\columnwidth]{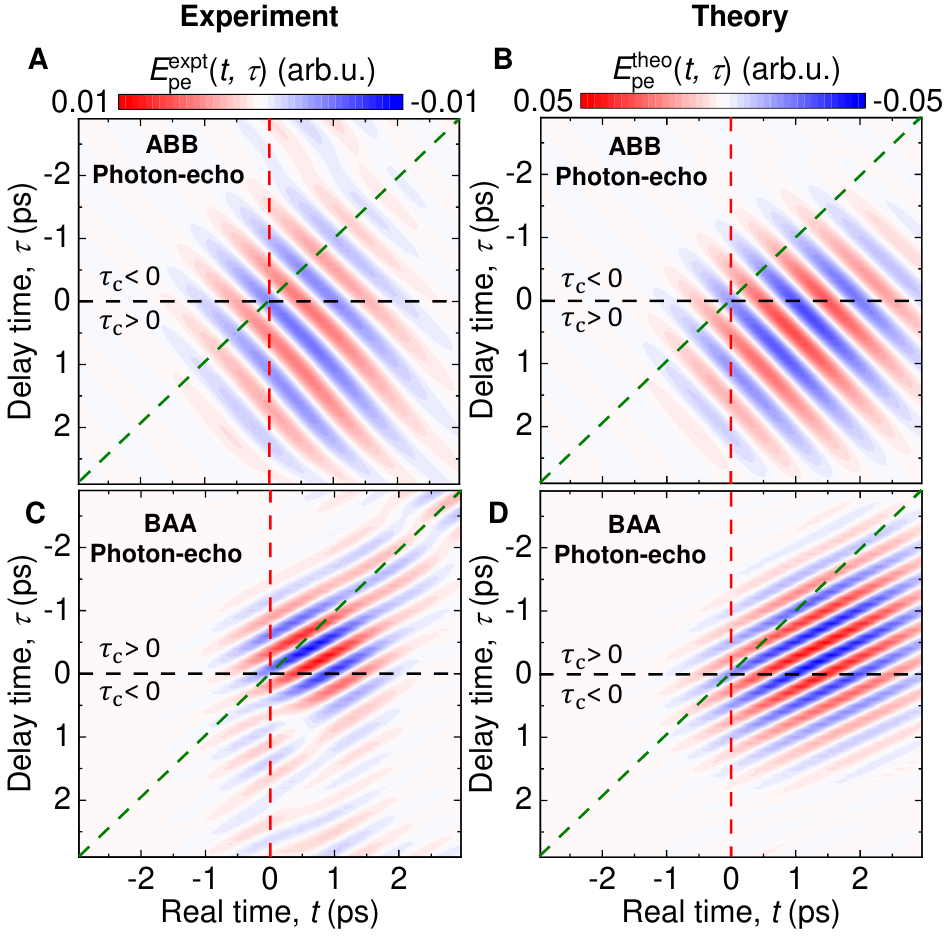}
	\caption{{\bf Temporal profiles of inverse Fourier-transformed photon-echo signals.} Contour plots of the ABB photon-echo signals obtained from {\bf (A)} experiment and {\bf (B)} theory. Contour plots of the ABB photon-echo signals obtained from {\bf (C)} experiment and {\bf (D)} theory. Green- and red-dashed lines trace the wavefronts of the two driving THz fields, while the black-dashed line marks zero delay ($\tau = 0$).}
	\label{fig5}
\end{figure}

\begin{figure}[t!]
	\centering
	\includegraphics[width=0.7\columnwidth]{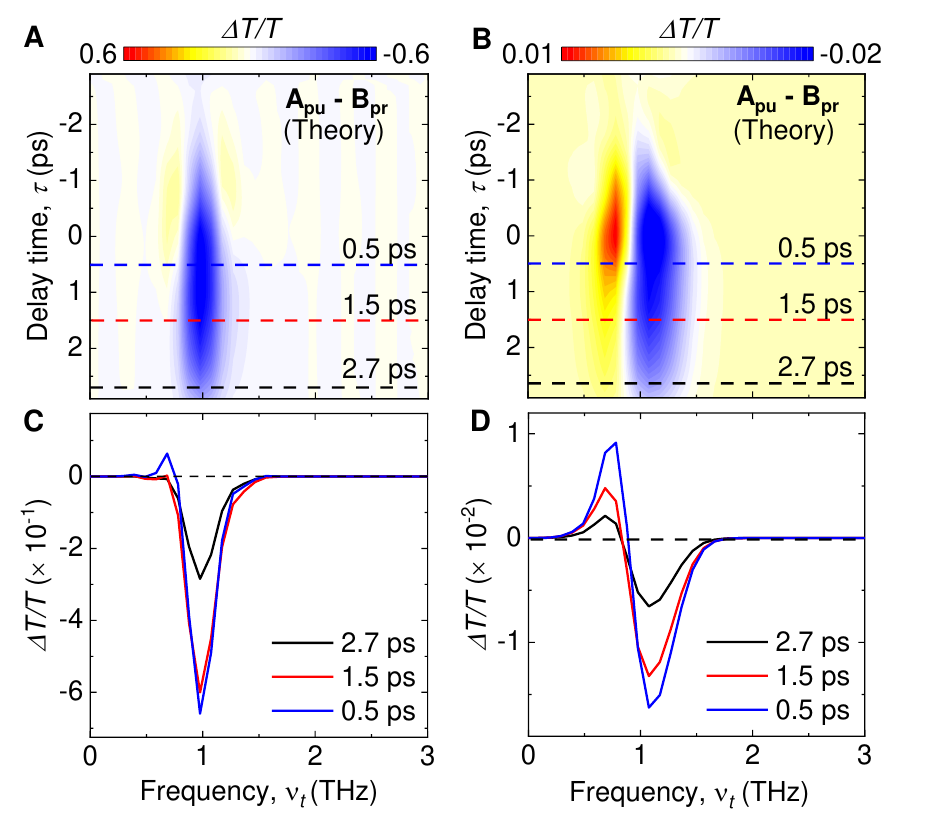}
	\caption{{\bf Simulated spectrally-resolved $\rm A_{\rm pu}-\rm B_{\rm pr}$ signals and line-cut  analysis.} {\bf (A, B)}  Contour plots of $\rm S^{AB}_{pp} (\nu_\textit{t},\tau)$ from theoretical modeling. In (B), the polarization is increased by an order of magnitude, which introduces a dispersive-like lineshape in the spectrally-resolved $\rm S^{AB}_{pp} (\nu_t, \tau)$ signal. {\bf (C, D)} Traces of $\rm S^{AB}_{pp} (\nu_\textit{t},\tau)$ obtained from (A) and (B) for three distinct delay times.}
	\label{fig6}
\end{figure}

The striking contrast between the ferroelectric and paraelectric phases of STO raises a fundamental question about the microscopic origin of the observed dynamics -- specifically, whether the nonlinear response retains signatures of strong local-field effects. Because photon-echo signals preserve the phase information corresponding to individual light-field interactions, they can be used to elucidate the role of local fields in the nonlinear optical response. Figure~\ref{fig5} displays the experimental and theoretically modeled contour plots of the ABB and BAA photon echo signals, demonstrating excellent qualitative agreement. Notably, both the measured and calculated profiles reveal a finite signal at negative coherence times ($\tau_{x}$). The emergence of photon echoes at the negative coherence times requires the system to preserve phase information from the preceding light-matter interactions. Within our microscopic theory, this coherence is mediated by the local field, which dynamically links successive THz field interactions. Thus, the observation of photon echoes at negative coherence times demonstrates that local fields strongly influence the material's macroscopic polarization even within the paraelectric phase, providing definitive evidence of local-field-driven coherence. To validate this mechanism and identify the non-perturbative threshold, we simulated the system by systematically increasing the spontaneous polarization by an order of magnitude. The resulting spectrally-resolved $\rm A_{\rm pu}-\rm B_{\rm pr}$ signal (Figure~\ref{fig6}B) and its corresponding time slices at 0.5\,ps, 1.5\,ps, and 2.7\,ps (Figure~\ref{fig6}D) exhibit a highly asymmetric, dispersive lineshape, confirming a transition into the non-perturbative regime of light-matter interaction. Such a dispersive spectral profile directly reflects a shift of the soft-mode resonance frequency, which serves as the experimental signature of the non-perturbative light-matter interaction in the soft-mode system.~\cite{Pal2021} Crucially, the coexistence of a photon-echo signal without an associated frequency shift isolates a distinct dynamic regime: local-field effects are strong enough to sustain coherent interactions and generate echoes at negative coherence times, but insufficient to modify the resonance frequency. This identifies a light-matter interaction regime wherein local-field-induced coherence persists independently of non-perturbative frequency shifts.

\section*{Conclusion}
In conclusion, we have elucidated the distinct roles of local-field effects and spontaneous polarization in driving a paraelectric STO system into the non-perturbative regime of light-matter interaction. Compressive in-plane strain successfully stabilized the system well away from its phase-transition boundary. Linear THz spectroscopy confirmed the hybrid nature of the identified soft mode by quantifying its electronic and ionic contributions. Although 2D THz pump-probe spectroscopy revealed no resonance frequency shifts -- confirming the system remains within the perturbative regime -- the photon-echo signals displayed robust local-field-induced coherence. This indicates that strong local fields alone cannot trigger non-perturbative dynamics. This behavior was validated by a theoretical model reproducing the key soft-mode nonlinearities. Ultimately, by systematically increasing the weight of spontaneous polarization in simulations, we recovered the expected non-perturbative behavior, thereby distinguishing the individual contributions of local fields and spontaneous polarization in engineered perovskite systems.

\section*{Methods}
\subsection{Growth and strain characterization:}
A 50-nm-thick, (001)-oriented epitaxial SrTiO$_3$ (STO) thin film was grown on a (001)-oriented (La$_{0.3}$Sr$_{0.7}$)(Al$_{0.65}$Ta$_{0.35}$)O$_3$ (LSAT) single-crystal substrate via pulsed laser deposition. Growth was performed at a substrate temperature of 700\,$^{\circ}$C under an O$_2$ partial pressure of 0.12\,mbar. A KrF excimer laser ($\lambda = 248$\,nm) was operated at a fluence of 1.14\,Jcm$^{-2}$ and a repetition rate of 2\,Hz. To minimize local strain relaxation, the sample was annealed {\it in situ} at 700\,$^{\circ}$C for 1\,h and subsequently cooled at 1\,K/min under the growth atmosphere. Reflection high-energy electron diffraction (RHEED) monitored the growth in real time, confirming high crystalline quality and layer-by-layer deposition, while the final thickness was verified via X-ray reflectivity (XRR). Both STO and LSAT exhibit cubic symmetry with bulk lattice constants of 3.905\,\text{\AA} and 3.868\,\text{\AA}, respectively, yielding a nominal in-plane compressive strain of $-0.95\%$. Furthermore, reciprocal space mapping (inset of Figure~\ref{fig1}(B)) demonstrates that the STO film is coherently strained, as evidenced by the alignment of the film and substrate peaks along the $Q_{\parallel}$ direction.

\subsection{Experimental details:}
To probe the soft mode in paraelectric STO, we performed linear terahertz time-domain spectroscopy (THz-TDS). The setup used a Ti:Sapphire laser operating at 800\,nm, with a 100\,fs pulse duration and a 1\,kHz repetition rate. The laser output was split, with 90\% of the beam power directed to generate THz radiation via optical rectification in a 0.5-mm-thick (110)-cut ZnTe crystal, while the remaining 10\% served as the sampling probe. The generated THz pulses were transmitted through the sample and collinearly focused alongside the probe beam onto a second (110)-oriented ZnTe detection crystal. The THz field-induced ellipticity of the probe beam was subsequently analyzed using a quarter-wave plate, a Wollaston prism, and a balanced photo-detector. To investigate the nonlinear THz response, we employed two-dimensional THz spectroscopy (2D-THz), where the relative arrival time between two collinear THz pulses was adjusted by a delay time $\tau$. Both THz transients were generated via optical rectification in (110)-cut ZnTe crystals. In our experiments, a combined THz electric field of $|E_{\rm AB}|=30$\,kV/cm was used The nonlinear response was isolated by subtracting the individual single-pulse traces from the combined two-pulse response, expressed as $E_{\rm NL} (t,\tau)= E_{\rm AB}(t,\tau)-E_{\rm A}(t,\tau)-E_{\rm B}(t)$. Because of the collinear excitation geometry, all nonlinear signals are encoded within a single 2D spectral map. All measurements were conducted at room temperature in an inert nitrogen environment to eliminate atmospheric water absorption. The temporal profiles of the transmitted THz pulses are provided in the supplementary material.

\subsection{Theoretical modeling:}
To model the microscopic origin of the nonlinear response, the theoretical simulations are performed using a self-consistent framework of coupled equations of motion. Here, we divide the entire system into two coupled subsystems: the electronic subsystem and the lattice vibrations. The lattice vibration is modeled as an ensemble of independent damped harmonic oscillators because the intrinsic anharmonicity corresponding to the phonon mode is extremely small. For every distinct lattice vibration indexed by $k$, the temporal evolution of the microscopic vibrational coherence ($p_{\rm vib, k}$) is given by the following equation of motion,
\begin{equation}\label{eq1}
\frac{\partial p_{{\rm vib},k}}{\partial t} =\left(i\,2\pi\nu_{{\rm vib},k}-\gamma_{{\rm vib},k}\right)p_{{\rm vib},k}+i\,\Omega_{{\rm vib},k}(t).
\end{equation}
Here, $\nu_{{\rm vib},k}$ denotes the fundamental resonance frequency, $\gamma_{{\rm vib},k}$ is the phenomenological damping rate. The vibrational Rabi frequency $\Omega_{{\rm vib},k}$ is given by,
\begin{equation}
\Omega_{{\rm vib},k}(t)=\frac{d_{{\rm vib},k} E_{{\rm local}}(t)}{\hslash}.
\end{equation}
The polarization introduced by the lattice vibration can be quantified as, 
\begin{equation}
P_{{\rm vib},k}=2N_{{\rm vib},k}\,d_{{\rm vib},k}\,{\rm Re}\!\left(p_{{\rm vib},k}\right),
\end{equation}
where $N_{{\rm vib},k}=(a_0b_0c_0)^{-1}$ is the density of a certain vibrational mode and $d_{{\rm vib},k}$ is the effective transition dipole moment corresponding to the lattice vibration. Here, $a_{0}$ = $b_{0}$ = $c_{0}$ are the lattice constants. In contrast to lattice vibrations, the electronic subsystem is intrinsically nonlinear and is modeled as an ensemble of two-level states governed by the density-matrix formalism. The temporal evolution of the electronic coherences ($\rho_{12}^{\rm el}$) and populations ($\rho_{22}^{\rm el}$) are given by:
\begin{equation}
\frac{\partial \rho_{12}^{{\rm el}}}{\partial t}=\left(i\,2\pi \nu_{\rm el}-\frac{1}{T_2^{\rm el}}\right)\rho_{12}^{\rm el}+i\,\Omega_{\rm el}(t)\left(1-2\rho_{22}^{\rm el}\right),\\
\end{equation}
\begin{equation}
\frac{\partial \rho_{22}^{\rm el}}{\partial t}=-\frac{1}{T_1^{\rm el}}\rho_{22}^{\rm el}+2\,\Omega_{\rm el}(t)\,{\rm Im}\!\left(\rho_{12}^{\rm el}\right).
\end{equation}
In these equations, we have phenomenologically incorporated the dephasing time ($T_2^{\rm el}$) and the population relaxation time ($T_1^{\rm el}$). The electronic Rabi frequency ($\Omega_{\rm el}$) is expressed as, 
\begin{equation}
\Omega_{\rm el}(t)=\frac{d_{\rm el}E_{\rm local}(t)}{\hslash},
\end{equation}
where $d_{\rm el}$ is the effective transition dipole moment corresponding to the electronic oscillators. The polarization corresponding to the electronic oscillator is quantified as, 
\begin{equation}
P_{\mathrm{el}}=2N_{\mathrm{el}}d_{\mathrm{el}}\,\mathrm{Re}\!\left(\rho_{12}^{\mathrm{el}}\right).
\end{equation}
Here, the saturation density of the electronic oscillators ($N_{\rm el}$) is approximately $a_{\rm ex}^{-3}$ ($a_{\rm ex}$ being the excitonic Bohr radius) such that $N_{{\rm vib},k} >> N_{\rm el}$. The driving terms ($\Omega_{{\rm vib},k},$ and $\Omega_{\rm el}$) for both subsystems are proportional to the local electric field ($E_{\rm local}$), which includes contributions from the externally applied THz field ($E_{\rm THz}$) and the Lorentz field caused by the induced polarization. The local electric field is thus given by:
\begin{equation}
E_{\rm local}=E_{\rm THz}+L(a_0,b_0,c_0)\left(P_{\rm el}+\sum_k P_{{\rm vib},k}\right).
\end{equation}
Since the system under study is paraelectric, i.e., cubic in nature (where the lattice constants are equal, $a_{0}=b_{0}=c_{0}$), we have taken the Lorentz factor ($L$) to be $1/3$. Note that the electronic transition dipoles are orders of magnitude larger than those of pure lattice excitations ($d_{\rm el}>>d_{{\rm vib},k}$). By solving the above-mentioned equations of motion in a self-consistent way, we extract $P_{\rm el}$ and $P_{\rm vib}$ for all delay times. Finally, the macroscopic re-emitted field ($E_{\rm em}$) is calculated directly from the time derivative of the total polarization as: 
\begin{equation}
E_{\rm em}=-\frac{\ell_{\rm sample}}{\varepsilon_0 c (n_{\rm sub}+1)}\left(\frac{\partial P_{\rm el}}{\partial t}+\sum_k\frac{\partial P_{{\rm vib},k}}{\partial t}\right).
\end{equation}
Here, $l_{\rm sample}$ is the thickness of the paraelectric layer and $n_{\rm sub}$ is the refractive index of the LSAT substrate. $\hslash, \varepsilon_0$, $c$ are the Planck's constant, vacuum permittivity, and velocity of light, respectively.

\clearpage

\clearpage
\begin{addendum}

\item[Lead contact:] Requests for further information and resources should be directed to and will be fulfilled by the lead contact, Shovon Pal (shovon.pal@niser.ac.in).

\item[Materials availability:] This study did not generate any new unique materials.

\item[Data and code availability:] The datasets analyzed in the current study are attached. Any additional data and the codes associated with the theoretical simulation supporting this study are available from the lead author upon reasonable request.

\item[Author Contributions:] All authors contributed to the discussion and interpretation of the experiment and to the completion of the manuscript. P.S. performed the linear and nonlinear THz experiments and the corresponding data analysis. I.F. and J.L., under the supervision of M.T., grew and characterized the samples. P.S. and K.V.G. developed the theoretical model, while P.S. analyzed the theoretical results. S.P. conceived the project and supervised the experiments. P.S. and S.P. drafted the manuscript.

\item[Acknowledgments:] P.S. and S.P. acknowledge the support from DAE through the projects RIN4001 and  RNI4011. S.P. also acknowledges the start-up support from DAE through NISER and SERB through SERB-SRG via Project No.~SRG/2022/000290. M.T. and I.E. acknowledge the Swiss National Science Foundation (SNSF) under projects no. 200021\_231428 and 200021-236413.

\item[Competing Interests:] The authors declare that they have no competing financial interests.

\item[Declaration of generative AI and AI-assisted technologies in the writing process:] During the preparation of the work, the authors used Gemini for spellchecking and grammar. After using the tool, the authors reviewed and edited the content as needed. The authors take full responsibility for the publication's content.

\end{addendum}


\clearpage
\begin{thebibliography}{1}
\bibitem{Venkataraman1979}
G. Venkataraman, Soft-modes and structural phase transitions, Bull. Mater. Sci. {\bf 1}, 3 (1979).

\bibitem{Scott1974}
J. F. Scott, Soft-mode spectroscopy: experimental studies of structural phase transitions, Rev. Mod. Phys. {\bf 46}, 83 (1974).

\bibitem{Petzelt2003}
J. Petzelt, P. Ku{\v{z}}el, I. Rychetsk{\`y}, A. Pashkin, and T. Ostapchuk, Dielectric response of soft modes in ferroelectric thin films, Ferroelectrics {\bf 288}, 169 (2003).

\bibitem{Kamba2020}
S. Kamba, Soft-mode spectroscopy of ferroelectrics and multiferroics: A review, APL Mater. {\bf 9}, 020704 (2020).

\bibitem{Barker1962}
A. S. Barker, JR. and M. Tinkham, Far-Infrared ferroelectric vibration Mode in SrTiO$_3$, Phys. Rev. {\bf 125}, 1527 (1962).

\bibitem{Cohen1992}
R. E. Cohen, Origin of ferroelectricity in perovskite oxides, Nature(London) {\bf 358}, 136 (1992).

\bibitem{Hannay1983}
J. H. Hannay, The Clausius-Mossotti equation: An alternative derivation, Eur. J. Phys. {\bf 4}, 141 (1983). 

\bibitem{Bucher1990}
M. Bucher, Review: Revaluation of the local field, J. Phys. Chem. Solids {\bf 51}, 1241 (1990).

\bibitem{Neb2026}
S. Neb, D. B. Shin, F. Burri, M. Hollm, E. W. de Vos, D. A. Kuznetsov, C. R. M\''{u}ller, A. Fedorov, S. A. Sato, A. Rubio, L. Gallmann, and U. Keller, Local fields reveal atomic-scale nonadiabatic carrier-phonon dynamics, Science, {\bf 391}, 75 (2026).

\bibitem{Maki1991}
J. J. Maki, M. S. Malcuit, J. E. Sipe, and R. W. Boyd, Linear and nonlinear optical measurements of the Lorentz local field, Phys. Rev. Lett. {\bf 67}, 972 (1991).

\bibitem{Pal2021}
S. Pal, N. Strkalj, C.-J. Yang, M. C. Weber, M. Trassin, M. Woerner, and M. Fiebig, Origin of terahertz soft-mode nonlinearities in ferroelectric perovskites, Phys. Rev. X {\bf 11}, 021023 (2021).

\bibitem{Folpini2017}
G. Folpini, K. Reimann, M. Woerner, T. Elsaesser, J. Hoja, and A. Tkatchenko, Strong local-field enhancement of the nonlinear soft-mode response in a molecular crystal, Phys. Rev. Lett. {\bf 119}, 097404 (2017).

\bibitem{Kozina2019}
M. Kozina, M Fechner, P. Marshik, T. van Driel, J. M. Glownia, C. Bernhard, M. Radovic, D. Zhu, S. Bonetti, U. Staub, and M. C. Hoffmann, Terahertz-driven phonon upconversion in $\rm SrTiO_3$, Nat. Phys. {\bf 15}, 387 (2019). 

\bibitem{Subedi2014}
A. Subedi, A. Cavalleri, and A. Georges, Theory of nonlinear phononics for coherent light control of solids, Phys. Rev. B {\bf 89}, 220301(R) (2014).

\bibitem{Yang2023}
C.-J. Yang, J. Li, M. Fiebig, and S. Pal, Terahertz control of many-body dynamics in quantum materials, Nat. Rev. Mater. {\bf 8}, 518 (2023).

\bibitem{Woerner2013}
M. Woerner, W. Kuehn, P. Bowlan, K. Reimann, and T. Elsaesser, Ultrafast two-dimensional terahertz spectroscopy of elementary excitations in solids, New J. Phys. {\bf 15},
025039 (2013).

\bibitem{Houver2019}
S. Houver, L. Huber, M. Savoini, E. Abreu, and S. L. Johnson, 2D THz spectroscopic investigation of ballistic
conduction-band electron dynamics in InSb, Opt. Express {\bf 27}, 10854 (2019).

\bibitem{Raab2019}
J. Raab, C. Lange, J. L. Boland, I. Laepple, M. Furthmeier, E. Dardanis, N. Dessmann, L. Li, E. H. Linfield, A. G.
Davies, M. S. Vitiello, and R. Huber, Ultrafast two-dimensional field spectroscopy of terahertz intersubband
saturable absorbers, Opt. Express {\bf 27}, 2248 (2019).

\bibitem{Markmann2021}
S. Markmann, M. Francki\'{e}, S. Pal, D. Stark, M. Beck, M. Fiebig, G. Scalari, and J. Faist, Two-Dimensional spectroscopy on a THz quantum cascade structure, Nanophotonics
{\bf 10}, 171 (2021).

\bibitem{Kuehn2011}
W. Kuehn, K. Reimann, M. Woerner, T. Elsaesser, and R. Hey, Two-dimensional terahertz correlation spectra of electronic excitations in semiconductor quantum wells, J. Phys. Chem. B {\bf 115}, 5448 (2011).

\bibitem{Somma2016prl}
C. Somma, G. Folpini, K. Reimann, M. Woerner, and T. Elsaesser, Two-phonon quantum coherences in indium
antimonide studied by nonlinear two-dimensional terahertz spectroscopy, Phys. Rev. Lett. {\bf 116}, 177401 (2016).

\bibitem{Somma2016} 
C. Somma, G. Folpini, K. Reimann, M. Woerner, and T. Elsaesser, Phase-resolved two-dimensional terahertz
spectroscopy including off-resonant interactions beyond the $\chi^3$ limit, J. Chem. Phys. {\bf 144}, 184202 (2016).

\bibitem{Peyghambarian1984}
N. Peyghambarian, H. M. Gibbs, J. L. Jewell, A. Antonetti, A. Migus, D. Hulin, and A. Mysyrowicz, Blue shift of the exciton resonance due to exciton-exciton interactions in a multiple-quantum-well structure, Phys. Rev. Lett. {\bf 53}, 2433 (1984).

\bibitem{Stone2009}
K. W. Stone, K. Gundogdu, D. B. Turner, X. Li, S. T. Cundiff, and K. A. Nelson, Two-quantum 2D FT electronic spectroscopy of biexcitons in GaAs quantum wells, Science {\bf 324}, 1169 (2009).

\bibitem{Gao2016}
F. Gao, S. T. Cundiff, and H. Li, Probing dipole-dipole interaction in a rubidium gas via double-quantum 2D spectroscopy, Opt. Lett. {\bf 41}, 2954 (2016).

\bibitem{Dolgaleva12012}
K. Dolgaleva1 and R. W. Boyd, Local-field effects in nanostructured photonic materials, Adv. Opt. Photonics. {\bf 4}, 1 (2012).

\bibitem{Katayama2012}
I. Katayama, H. Aoki, J. Takeda, H. Shimosato, M. Ashida, R. Kinjo, I. Kawayama, M. Tonouchi, M. Nagai, and K. Tanaka, Phys. Rev. Lett. {\bf 108}, 097401 (2012).

\bibitem{Efe2024}
I. Efe, B. Yan, M. Trassin, Engineering of ferroelectricity in thin films using lattice chemistry: A perspective, Appl. Phys. Lett. {\bf 125}, 150503 (2024).

\bibitem{Knoester1989}
J. Knoester and S. Mukamel, Nonlinear Optics Using the Multipolar Hamiltonian: The Bloch-Maxwell equations and local fields, Phys. Rev. A {\bf 39}, 1899 (1989).

\bibitem{Schmitt-Rink1991}
S. Schmitt-Rink, S. Mukamel, K. Leo, J. Shah, and D. S. Chemla, Stochastic theory of time-resolved four-wave mixing in interacting media, Phys. Rev. A {\bf 44}, 2124 (1991).

\bibitem{Haeni2004}
J. H. Haeni, P. Irvin, W. Chang, R. Uecker, P. Reiche, Y. L. Li, S. Choudhury, W. Tian, M. E. Hawley, B. Craigo, A. K. Tagantsev, X. Q. Pan, S. K. Streiffer, L. Q. Chen, S. W. Kirchoefer, J. Levy, and D. G. Schlom, Room-temperature ferroelectricity in strained SrTiO$_3$, Nature {\bf 430}, 758 (2004).

\bibitem{Katayama2008}
I. Katayama, H. Shimosato, D. S. Rana, I. Kawayama, M. Tonouchi, and M. Ashida, Hardening of the ferroelectric soft mode in SrTiO$_3$ thin films, Appl. Phys. Lett. {\bf 93}, 132903 (2008).

\bibitem{Sirenko2000}
A. A. Sirenko, C. Bernhard, A. Golnik, A. M. Clark, J. Hao, W. Si, and X. X. Xi, Soft-Mode hardening in SrTiO$_3$ thin films, Nature (London) {\bf 404}, 373 (2000).

\bibitem{Tu2003}
J. J. Tu, C. C. Homes, and M. Strongin, Optical properties of ultrathin films: evidence for a dielectric anomaly at the insulator-to-metal transition, Phys. Rev. Lett. {\bf 90}, 017402
(2003).

\bibitem{Cochran1960}
W. Cochran, Crystal stability and the theory of ferroelectricity, Adv. Phys. {\bf9}, 387 (1960).

\bibitem{Dutta2025}
A. Dutta, C. Tzschaschel, D. Priyadarshi, K. Mikuni, T. Satoh, R. Mondal, and S. Pal, Evidence of relativistic field-derivative torque in nonlinear THz response of magnetization dynamics, Adv. Funct. Mater. {\bf35}, 2414582 (2025). 

\bibitem{Dutta2025JPCM}
A. Dutta, P. Shee, A. Halder, and S. Pal, 2D THz spectroscopy: exploring the nonlinear dynamics in quantum materials, J. Phys.: Condens. Matter {\bf 37}, 203002 (2025).

\bibitem{Hamm2011}
P. Hamm, and M. Zanni, Concepts and methods of 2D infrared spectroscopy (University press, Cambridge, England, 2011).

\bibitem{Elsaesser2019}
T. Elsaesser, K. Reimann, and M. Woerner, Concepts and applications of nonlinear terahertz spectroscopy, IOP concise physics (A Morgan and Claypool publication, Sun Rafael, 2019).

\bibitem{Mukamel1995}
S. Mukamel, Principles of nonlinear optical spectroscopy, (Oxford University Press, New York, 1995).

\bibitem{Huang2026}
C. Huang, M. Mootz, L. Luo, I. E. Perakis, and J. Wang, Terahertz 2D coherent spectroscopy for probing and controlling multicorrelations in quantum matter, Nat. Rev. Phys. {\bf 8}, 171 (2026).

\bibitem{Haldar2026}
A. Haldar, S. Prabhu, S. Pal, Unveiling nonlinearities of electromagnetically induced transparency in a THz metamaterial, Adv. Funct. Mater. {\bf 36}, e75422 (2026).



\end{thebibliography}
\end{document}


\maketitle
\begin{affiliations}
\item School of Physical Sciences, National Institute of Science Education and Research, Jatni, 752 050 Odisha, India
\item Homi Bhabha National Institute, Training School Complex, Anushakti Nagar, 400 094 Mumbai, India
\item Department of Materials, ETH Zurich, 8093 Zurich, Switzerland
\end{affiliations}


\section{Relative electronic and ionic contribution to the soft mode}

 \begin{figure}[t!]
	\centering
	\includegraphics[width=1.0\columnwidth]{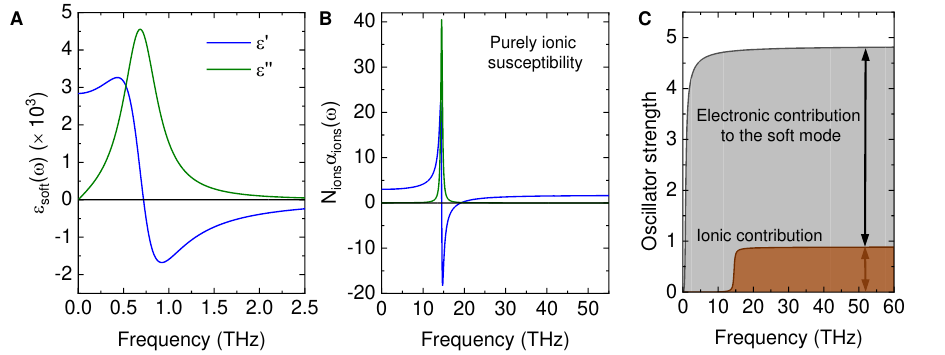}
	\caption{{\bf Relative electronic and ionic contribution} {\bf (A)} Real and imaginary parts of the soft-mode dielectric function calculated from the Lorentzian model. {\bf (B)} Real and imaginary parts of the ionic susceptibility excluding local field contribution. {\bf (C)} Integral (or equivalently the oscillator strength) of the real part of the conductivities corresponding to the electronic and ionic oscillators.}
	\label{fig3}
\end{figure}

 \begin{figure}[t!]
	\centering
	\includegraphics[width=1.0\columnwidth]{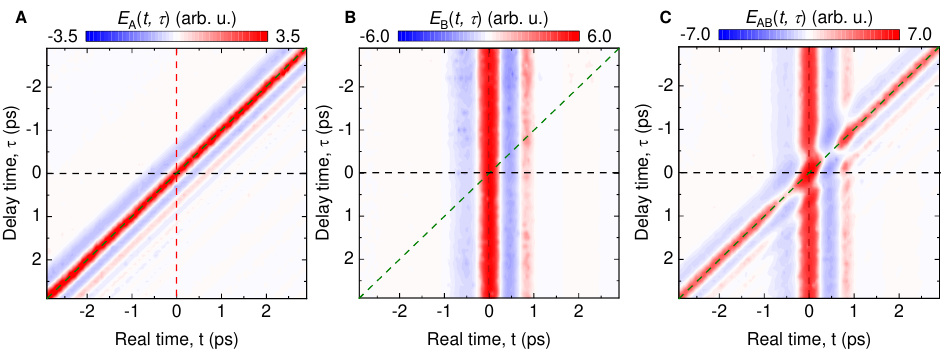}
	\caption{{\bf Single pulse and both pulse response transmitted through the sample} {\bf (A, B)} Contour plots of individual THz pulse, $E_{\rm A}(t,\tau)$ and ($E_{\rm B}(t,\tau)$, transmitted through the sample. {\bf (C)} Contour plots of both THz pulses, $E_{\rm AB}(t,\tau)$, transmitted simultaneously through the sample. The green- and red-dashed lines indicate the propagation wavefront of the THz driving fields A and B, respectively. The black-dashed line denotes the zero-delay time.}
	\label{fig3}
\end{figure}

We investigate the influence of electronic and lattice degrees of freedom on the soft mode resonance and contrast it with the behavior reported for its ferroelectric counterpart~\cite{Pal2021}. To quantify the relative contributions of electronic and ionic components to the linear THz response, we fitted the complex dielectric function (shown in Figure~1E of the main manuscript) using a damped harmonic oscillator model given by, 
\begin{equation}\label{eq1}
\begin{split}
    \epsilon_{\rm soft}(\omega) =\epsilon_{\infty}+\frac{\omega_{\rm ions}^2}{\omega_{\rm soft}^2-i\omega\gamma_{\rm soft}-\omega^2},    
\end{split}
\end{equation}
where $\omega_{\rm soft}$ is the soft mode frequency, $\gamma_{\rm soft}$ is the damping rate and $\omega_ {\rm ions} = Z_{\rm eff}^2 \rm N/\epsilon_{0}\rm M_{\rm red}$ is the ionic plasma frequency. Here, $Z_{\rm eff}$ is the Born effective charge, N is the density of the soft mode oscillators, $\epsilon_0$ is the vacuum permittivity, and $\rm M_{\rm red} = [1/(\rm M_{\rm Sr}+\rm M_{\rm Ti})+1/\rm 3M_{\rm O}]^{-1}$ is the reduced mass.

Considering only the ionic contribution, the fitting procedure yields an effective charge of approximately $Z_{\rm eff}\approx 10 \rm e_0$ ($\rm e_0$ is the elementary charge of the electron), which is nonphysical to reconcile with the realistic charge on the ions, indicating that the soft mode has a hybrid nature. A similar discrepancy was also observed in studies of the ferroelectric STO, where purely phononic models yielded unrealistically large effective charges~\cite{Pal2021}. This apparent inconsistency can be solved by introducing the local field effects. Within the framework of the Clausius-Mossotti relation, the macroscopic dielectric response corresponding to the ions redistributes and yields a physically reasonable value of the effective charge.

Based on Cochran's equations, we have further evaluated the relative electronic and ionic contribution by analyzing the spectral weight of the real part of conductivity \cite{Cochran1960}. Our analysis reveals that the electronic contribution to the soft mode is approximately 5 times higher than the purely ionic contribution, even in the paraelectric phase, which is consistent with the results obtained in the study of the ferroelectric STO~\cite{Pal2021}. This shows that the hybrid nature of the soft mode in STO persists regardless of whether it is ferroelectric or paraelectric. Thus, the strong electronic contributions and the incorporation of the local-field effects resolve the apparent discrepancy in the enhanced value of $\rm Z_{\rm eff}$. This naturally raises the question of how these strong electronic contributions and local-field effects influence the nonlinear response of a paraelectric material far from the ferroelectric phase transition.